%% file: chiir23-generative-model-as-index-frame.tex
\documentclass[sigconf]{acmart} 
\setcopyright{rightsretained}

\usepackage{chiir23-generative-model-as-index-frame}
\graphicspath{{figures/}}

\begin{document}

\input{chiir23-generative-model-as-index-pre}

\input{chiir23-generative-model-as-index-part1}
\input{chiir23-generative-model-as-index-part2}
\input{chiir23-generative-model-as-index-part3}
\input{chiir23-generative-model-as-index-part4}
\input{chiir23-generative-model-as-index-part5}
\input{chiir23-generative-model-as-index-sum}

\begin{raggedright}
\bibliographystyle{ACM-Reference-Format}
\bibliography{chiir23-generative-model-as-index-lit}
\end{raggedright}

\end{document}

%% file: chiir23-generative-model-as-index-pre.tex
\title[The Infinite Index: Information Retrieval on Generative Text-To-Image Models]{The Infinite Index:\texorpdfstring{\\}{ }Information Retrieval on Generative Text-To-Image Models}

\author{Niklas Deckers}
\orcid{0000-0001-6803-1223}
\affiliation{%
  \institution{Leipzig University and ScaDS.AI}
  \city{}
  \country{}
}
\email{}

\author{Maik Fröbe}
\orcid{0000-0002-1003-981X}
\affiliation{%
  \institution{Friedrich-Schiller-Universität Jena}
  \city{}
  \country{}
}
\email{}

\author{Johannes Kiesel}
\orcid{0000-0002-1617-6508}
\affiliation{%
  \institution{Bauhaus-Universität Weimar}
  \city{}
  \country{}
}
\email{}

\author{Gianluca Pandolfo}
\orcid{0000-0002-7656-0560}
\affiliation{%
  \institution{Bauhaus-Universität Weimar}
  \city{}
  \country{}
}
\email{}

\author{Christopher Schröder}
\orcid{0000-0002-7081-8495}
\affiliation{%
  \institution{Leipzig University}
  \city{}
  \country{}
}
\email{}

\author{Benno Stein}
\orcid{0000-0001-9033-2217}
\affiliation{%
  \institution{Bauhaus-Universität Weimar}
  \city{}
  \country{}
}
\email{}

\author{Martin Potthast}
\orcid{0000-0003-2451-0665}
\affiliation{%
  \institution{Leipzig University and ScaDS.AI}
  \city{}
  \country{}
}
\email{}

\begin{abstract}
Conditional generative models such as DALL-E and Stable Diffusion generate images based on a user-defined text, the prompt. Finding and refining prompts that produce a desired image has become the art of prompt engineering. Generative models do not provide a built-in retrieval model for a user's information need expressed through prompts. In light of an extensive literature review, we reframe prompt engineering for generative models as interactive text-based retrieval on a novel kind of ``infinite index''. We apply these insights for the first time in a case study on image generation for game design with an expert. Finally, we envision how active learning may help to guide the retrieval of generated images.
\end{abstract}

\keywords{case study, evaluation, generative models, image retrieval}

\ccsdesc[500]{Information systems~Search engine indexing}
\ccsdesc[500]{Information systems~Users and interactive retrieval}
\ccsdesc[500]{Information systems~Image search}
\ccsdesc[500]{Information systems~Novelty in information retrieval}
\ccsdesc[500]{Information systems~Search engine architectures and scalability}
\ccsdesc[500]{Information systems~Users and interactive retrieval}
\ccsdesc[500]{Information systems~Image search}

\makeatletter
\def\@copyrightowner{}
\def\@copyrightpermission{\vspace{\baselineskip}}
\@acmownedfalse
\makeatother
\copyrightyear{}
\def\copyright{\href{http://creativecommons.org/licenses/by/4.0/}{\includegraphics{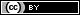}}\\\href{http://creativecommons.org/licenses/by/4.0/}{This work is licensed under a Creative Commons Attribution International 4.0 License.}}

\acmYear{2023}
\acmConference[CHIIR '23]{ACM SIGIR Conference on Human Information
Interaction and Retrieval}{March 19--23, 2023}{Austin, TX, USA}
\acmBooktitle{ACM SIGIR Conference on Human Information Interaction
and Retrieval (CHIIR '23), March 19--23, 2023, Austin, TX, USA}
\acmDOI{10.1145/3576840.3578327}
\acmISBN{979-8-4007-0035-4/23/03}

\maketitle

%% file: chiir23-generative-model-as-index-part1.tex
\section{Introduction}

Conditional generative models allow the generation of a desired output based on a user-specified condition. For generative text-to-image models such as DALL-E~\cite{ramesh:2021} or Stable Diffusion~\cite{rombach:2022}, this means that the model generates images conditional on a text description known as a prompt. For a user, the prompt is the primary means of controlling the generated image. If an ad hoc prompt does not produce a satisfactory result, the user usually interacts with the model by adjusting the prompt until they get one, or they give up after a few tries. Since such systematic refinement of prompts is often necessary to achieve a satisfactory result, writing prompts has evolved into the art of {\em prompt engineering}~\cite{reynolds:2021,liu:2022,oppenlaender:2022}, for which users exchange best practices in new communities. But even using examples from others, it's often not obvious how to change a prompt to steer image generation in a particular direction.

As a new perspective on the use of conditional generative models in general, we interpret them as a search engine index. Under this interpretation, the prompt is a request that represents a user's need for information. Prompt engineering can then be considered a form of interactive text-based retrieval, in which a user interacts with the model by modifying their prompt as if to refine their query to find a result that meets their needs. This raises a number of new challenges: When using a generative model, the initiative currently lies solely with the user, without support from the model as a ``retrieval system''. There is no intermediary retrieval model to help users produce satisfactory images fast(er), if not ad hoc. The manual refinement of prompts is not supported by system-side log analysis and query expansion. There is no operationalization of the concept of image relevance, which is needed for ranking images, and thus essential when many images are generated.

A striking difference from traditional retrieval is that when generative models are used as an index, new results are generated rather than existing ones retrieved.%
\footnote{Generative model occasionally reproduce parts of their training data~\cite{lee:2022,somepalli:2022}.}
A non-empty result is returned for every conceivable query prompt. This includes query prompts for which a traditional retrieval system would return no results. Also, the number of different results that can be generated per query prompt is not conceptually limited, but only by the available computational capacity for model inference. Thus, a generative model is effectively an ``infinite index''.

Our contribution is to explore this perspective on generative models as indexes in four ways, focusing on text-to-image generation:
\Ni
Section~\ref{part2} presents a literature survey on image generation, text-based image retrieval, retrieval for creative tasks, and interactive retrieval.
\Nii
Section~\ref{part3} conceptualizes generative text-to-image models as an index integrated into a retrieval system: from the user perspective, the query language and interaction methods are presented, and from the system perspective, retrieval technologies capable of supporting retrieval are examined. Requirements for the evaluation of retrieval systems based on generative models are also presented.
\Niii
Based on these findings, Section~\ref{part4} presents a case study of image generation. For creative tasks in game design, we observe an expert and highlight several issues related to currently available technology.
\Niv
Finally, based on the insights gained, Section~\ref{part5} discusses an active learning approach to interactive retrieval to guide image generation using generative models. 

%% file: chiir23-generative-model-as-index-part2.tex
\section{Background and Survey}
\label{part2}

We review the relevant literature to place retrieval on generative models in the context of established concepts.

\subsection{Image Generation}

In image synthesis, \citet{brock:2018} and \citet{goodfellow:2020} have achieved promising results with generative adversarial networks~(GANs) that allow images to be generated from the distribution of given training images. Autoregressive transformer models as per \citet{razavi:2019} and \citet{ramesh:2021} have proven to be effective for high-resolution image synthesis. \citet{dhariwal:2021} has recently shown that diffusion models~\cite{sohl:2015} are capable of outperforming traditional models such as~GANs in image synthesis. In addition, \citet{rombach:2022} have shown how to condition the generated images on text. This forms the basis for text-to-image models, which are often trained on datasets of text--image pairs~\cite{schuhmann:2021}.

Table~\ref{table-image-generation-models} provides an overview of relevant text-to-image models, starting with diffusion models such as DALL$\cdot$E by \citet{ramesh:2021} and Imagen by \citet{saharia:2022}. Most models are only accessible via a web interface. Their code and model weights are not publicly available. Stable Diffusion by \citet{rombach:2022} achieved great impact not only because of its impressive results, but also because the model itself was made publicly available. As a result, it was rapidly adapted and now serves as the basis for numerous new applications. More recent approaches pursue other research goals: eDiff-I by \citet{balaji:2022} introduces an ensemble of export-denoising networks that allow different behavior at different noise levels. This increases the number of parameters, but also improves the results. Muse by \citet{chang:2023} uses a discrete token space instead of a pixel space to increase efficiency.

\input{table-image-generation-models}

\subsection{Image Retrieval}

While text-to-image models are relatively new, image retrieval has a long history of research. Two cases are distinguished in the literature: In content-based image retrieval, the user enters an image as a query, while in text-based image retrieval, the user makes a textual query. Content-based image retrieval systems aim to bridge the gap between the semantic meaning of images and their quantified visual features through sophisticated image representations~\cite{li:2021}. Once a collection of images is represented and indexed, the representation of the query image is used for similarity-based search and ranking. Text-based image retrieval has often focused on retrieval based on image metadata and tags in the past, which is why it is sometimes referred to as annotation-based, concept-based, or keyword-based image retrieval. Some approaches also generate textual representations for unannotated images, e.g., using optical character recognition~\cite{unar:2019}, clustering images with and without annotations~\cite{lin:2004}, or using image captioning methods~\cite{iyer:2017}.

Some studies have examined users' search interactions with a text-based image retrieval system. \citet{choi:2013} analyzed the search logs of 29~students and found that participants changed their textual queries more frequently to refine their results. \citet{hollink:2011} studied the image search behavior of news professionals and showed that they often modified their queries by following semantic relationships of query terms, e.g., searching first for images about a person and then for images about their spouse.

\citet{cho:2021} took a closer look at why people search for images. In their study of 69~papers, they identified seven information need categories
\Ni
entertainment,
\Nii
illustrations (explanation or clarification of details, e.g., creating presentation slides or preparing study material),
\Niii
images for aesthetic appreciation (e.g., for desktop backgrounds),
\Niv
knowledge construction (four sub-categories: information processing, information dissemination, learning, and ideation),
\Nv
eye-catchers (e.g., to grab audiences' attention),
\Nvi
inspiring images, and 
\Nvii
images for social interactions (e.g., images to trigger emotions).
They also found seven categories of problems that could affect a user's ability to find the images they were looking for:
\Na
semantic issues, i.e., related to employed terminology,
\Nb
content-based issues, i.e., related to describing content of images,
\Nc
technical limitations of retrieval systems,
\Nd
lacking aboutness or relevance of retrieved images,
\Ne
lacking inclusivity with regard to cultural or linguistic aspects of the user,
\Nf
lacking skills in handling search technology, and
\Ng
cognitive overload.
As we discuss in Section~\ref{part3}, most of these requirements and issues are also relevant to retrieval from text-to-image models.

\subsection{User Feedback for Image Generation}

\enlargethispage{\baselineskip}
Based on GANs, \citet{ukkonen:2020} have proposed and implemented systems for relevance feedback and \citet{yangliu:2022} for exploratory search. This was to overcome the lack of prompts in GANs to condition image generation, leaving users with little control over the generated images. Similar techniques to incorporate relevance feedback could be considered for text-to-image models.

\subsection{Retrieval for Creative Tasks}

Text-to-image models are particularly suited to artistic and creative applications, raising the question of whether there are parallels between such applications and the literature on creative task search. Interestingly, text-to-image models have quickly led to the formation of communities dedicated not only to the use of these tools, but also to prompt engineering and the sharing of successful image generation techniques.%
\footnote{The Midjourney Discord server has more than 8~million members (as of January~2023).}
This development is consistent with the formation of creative communities by artists in other art genres~\cite{hemmig:2008}. On the other hand, such strong community building is somewhat surprising, since artisans generally rely less on human sources~\cite{lee:2019}.

Several studies have already specifically analyzed user behavior and goals in creative tasks. \citet{chavula:2022} investigated the information behavior of 15~graduate students in creative web search tasks using questionnaires and the think-aloud method. They identified four creative thinking processes that participants switched back and forth between: planning creative search tasks (i.e., deciding on a vague idea), searching for new ideas, synthesizing search results, and organizing ideas. \citet{palani:2021} use log analyses and self-reports in a study of 34~design students. They observed three main goals of the students: To get an overview of the information space, to discover design patterns and criteria, and to get inspired and develop ideas. In the study, special attention was paid to the fact that participants initially had difficulty finding appropriate terms to describe their information needs, but then arrived at appropriate terms by quickly querying and reformulating queries. They also note that participants typically go through a divergent exploration phase before a convergent synthesis phase. Based on a previous online survey and study~\cite{zhang:2019,zhang:2020}, \citet{li:2022} examine the information behavior in a diary study of 11~university students on self-selected creative tasks. They use Sawyer's eight-step creativity framework~\cite{sawyer:2012} and focus specifically on the use of information resources (search, images, Q\&A, social sites, videos). They grouped them into five categories: Searching for specific information, supporting creative processes, learning definitional domains, learning procedural knowledge, and managing (organizing) found information. Especially with images, they distinguish specific uses (e.g., as on Pinterest, Instagram, Tumblr, Flickr, and image search): Support ideation and other creative processes, see finished examples, find out what one likes or dislikes, and manage and overview found information. They found that image search engines were primarily used to search for a wide range of images, while image sites like Pinterest and Instagram were often used to search for high-quality images by specific artists or professionals. In summary, we identify three common topics when searching for creative tasks: Searching to learn, to get inspired, and to get an overview. We also observe these behaviors in our case study (Section~\ref{part4}).

\subsection{Interactive Retrieval}

\enlargethispage{\baselineskip}
Interactive retrieval explores users' information behavior during and beyond search, as well as the development of new interaction methods to assist them~\cite{ruthven:2008}.  In relation to our work, we review relevant research on query understanding based on query logs as a source of user interaction data.

\paragraph{Query Log Analysis}
\citet{joachims:2007} introduced query log analysis for web search, which has since become a valuable tool, e.g., for improving retrieval effectiveness and studying user behavior~\cite{broder:2002,jansen:2005,jansen:2007,jansen:2009}. \citet{broder:2002}, for example, established a taxonomy for web search queries showing that web search queries are divided into informational, navigational, and transactional queries, which is still the case today~\cite{alexander:2022}. A further categorization derives from \citeauthor{jansen:2009}'s~\cite{jansen:2009} work on query reformulation: queries are either generalizations (subset of words), specializations (superset of words), synonyms, or other topics. Today, query logs are used for creating large training datasets for retrieval models based on transformers~\cite{rekabsaz:2021,nguyen:2016} and remain an important asset. 

\paragraph{Query Reformulation}
Query reformulation approaches aim to improve the effectiveness of retrieval by replacing the original query with substituted or extended reformulations~\cite{dang:2010}. Here, the reformulation of a query can be either precision-oriented (when a term is replaced by a more specific one) or recall-oriented (when the query is expanded). \citet{jansen:2005} shows that searchers do not start with perfect queries but reformulate them instead: more than 50\% of searchers reformulate at least one query during a search. Approaches to automatic query expansion, such as RM3~\cite{jalee:2004}, can use (pseudo) relevance feedback to add new (weighted) terms to the original query, thus solving the vocabulary mismatch problem that occurs in text retrieval. However, it is not yet clear which reformulations are helpful in which situations when working creatively with generative text-to-image models (i.e., precision-oriented or recall-oriented reformulations).

\paragraph{Query Suggestion}
Search engines assist their users and offer a list of suggested queries for an input query~\cite{bhatia:2011}, which is called query auto-completion~\cite{cai:2016} if the query is incomplete. Query suggestions are important; according to \citet{feuer:2007}, 30\%~of queries in a commercial query log are suggested to users beforehand. Likewise, \citet{cucerzan:2004} notes that spelling corrections are required for 10-15\%~of queries with spelling errors. In addition, query suggestions often aim to assist users by displaying related terms~\cite{huang:2003}, where \citeauthor{jansen:2007}'s~\cite{jansen:2007} analysis shows that suggested related terms are also heavily used. However, it is important not to overwhelm users and rather show fewer alternatives for suggestions than many~\cite{white:2007}. Overall, users value the interaction methods used in ``traditional'' search engines, and we believe that offering similar ones for retrieval interfaces built on generative text-to-image models will provide benefits to users with creative tasks.

%% file: table-image-generation-models.tex
\begin{table*}[!t]
\small
\centering
\fontsize{9pt}{10pt}\selectfont%
\renewcommand{\tabcolsep}{3.7pt}
\renewcommand{\arraystretch}{1.1}

\newcommand{\bslink}[1]{\href{#1}{\includegraphics[scale=0.33]{external-link}}}

\caption{Overview of the most relevant text-to-image models (\includegraphics[scale=0.27]{external-link-black}~web link; $^*$\,replicated; $^\dagger$\,includes the text encoder).}
\label{table-image-generation-models}

\begin{tabular}{@{}lrrlcccllc@{}}
\toprule
  \multicolumn{2}{@{}c@{}}{\textbf{Text-to-image model}} & \multicolumn{2}{@{}c@{}}{\textbf{Training data}}                                                                                                        & \multicolumn{3}{@{}c@{}}{\textbf{Open Source}}                                                                                                                                                       & \multicolumn{3}{@{}c@{}}{\textbf{Reference}}                                                     \\
\cmidrule(){1-2}\cmidrule(l@{\tabcolsep}r@{\tabcolsep}){3-4}\cmidrule(l@{\tabcolsep}r@{\tabcolsep}){5-7}\cmidrule(l@{\tabcolsep}){8-10}
  Name             &              \kern-2.5em Parameters &              Size & Source                                                                                                                              &                                  Code                                  &                             Data                             &                                 Model                                 & Publication            & \kern-1em Link                                           & Month / Year \\
\midrule
  DALL$\cdot$E     &                               12\,B & {\color{gray}n/a} & Custom web crawl~\bslink{https://github.com/openai/DALL-E/blob/master/model_card.md\#training-data}                                 & \bslink{https://github.com/lucidrains/DALLE-pytorch}\,$^*$\kern-0.5em  &                              --                              & \bslink{https://github.com/robvanvolt/DALLE-models}\,$^*$\kern-0.5em  & \citet{ramesh:2021}    & \bslink{https://openai.com/blog/dall-e/}                 &  01 / 2021   \\
  DALL$\cdot$E~2   &                              3.5\,B & {\color{gray}n/a} & Custom web crawl, licensed sources~\bslink{https://github.com/openai/dalle-2-preview/blob/main/system-card.md\#model-training-data} & \bslink{https://github.com/lucidrains/DALLE2-pytorch}\,$^*$\kern-0.5em &                              --                              & \bslink{https://huggingface.co/laion/DALLE2-PyTorch}\,$^*$\kern-0.5em & \citet{ramesh:2022}    & \bslink{https://openai.com/dall-e-2/}                    &  04 / 2022   \\
  Imagen           &                              4.6\,B &            860\,M & 400\,M \cite{schuhmann:2021} from Common Crawl                                                                                      & \bslink{https://github.com/lucidrains/imagen-pytorch}\,$^*$\kern-0.5em &                              --                              &  \bslink{https://github.com/cene555/Imagen-pytorch}\,$^*$\kern-0.5em  & \citet{saharia:2022}   & \bslink{https://imagen.research.google/}                 &  05 / 2022   \\
  Midjourney       &                   {\color{gray}n/a} & {\color{gray}n/a} & {\color{gray}n/a}                                                                                                                   &                                   --                                   &                              --                              &                                  --                                   & \citet{salkowitz:2022} & \bslink{https://midjourney.com/}                         &  07 / 2022   \\
  Stable Diffusion &                              0.9\,B &            400\,M & Common Crawl; cf.~\citet{schuhmann:2021}                                                                                            &        \bslink{https://github.com/Stability-AI/stablediffusion}        & \bslink{https://huggingface.co/datasets/laion/laion5B-index} &             \bslink{https://huggingface.co/stabilityai/}              & \citet{rombach:2022}   & \bslink{https://github.com/Stability-AI/stablediffusion} &  08 / 2022   \\
  eDiff-I          &         9.1\,B$^\dagger$\kern-0.5em & {\color{gray}n/a} & {\color{gray}n/a}                                                                                                                   &                                   --                                   &                              --                              &                                  --                                   & \citet{balaji:2022}    & \bslink{https://deepimagination.cc/eDiff-I/}             &  11 / 2022   \\
  Muse             &                                3\,B &            460\,M & {\color{gray}n/a}; cf.~\citet{saharia:2022}                                                                                         &                                   --                                   &                              --                              &                                  --                                   & \citet{chang:2023}     & \bslink{https://muse-model.github.io/}                   &  01 / 2023   \\
\bottomrule
\end{tabular}
\end{table*}

%% file: chiir23-generative-model-as-index-part3.tex
\begin{figure*}
\centering
\includegraphics{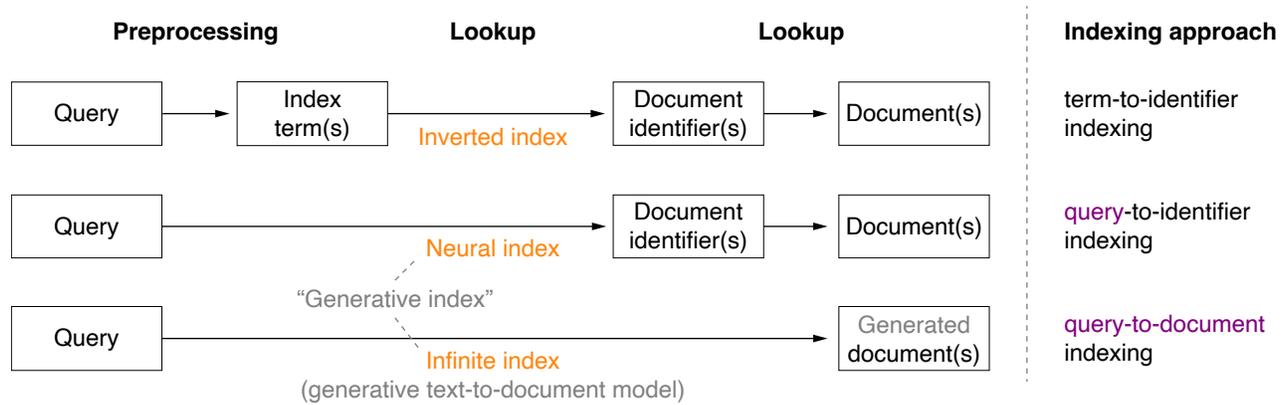}
\caption{Overview of indexing approaches in information retrieval. The top row shows the classic term-to-identifier indexing approach, the middle rows the recent query-to-identifier indexing approach, and the bottom row the new query-to-document indexing approach introduced in this paper.}
\label{index-technologies}
\Description{The images shows three flowcharts: The first one shows the classic term-to-identifier indexing approach, where queries are preprocessed to index terms. The index terms are looked up in an inverted index to obtain document identifiers. The document identifiers are used to lookup the actual documents in a document store. The second flowchart shows the query-to-identifier indexing approach, where queries are mapped to document identifiers which are then used to lookup documents in the document store. The third one is the new query-to-document indexing approach, where queries are directly mapped to generated documents.}
\end{figure*}

\section{Text-To-Image Generation as Search}
\label{part3}

Considering a text-to-image model as a virtually infinite index, a prompt as a query, and prompt engineering as a form of user-driven query refinement yields a rudimentary retrieval system (Section~\ref{infinite-index-in-ir}). In the following, the interaction methods (Section~\ref{interaction-methods}) that are (potentially) available to users and the retrieval technologies (Section~\ref{missing-intermediary}) that are (potentially) applicable to such a retrieval system are examined in detail. Subsequently, requirements for the evaluation of such a system are formulated (Section~\ref{ir-evaluation}).

\subsection{Classification of the ``Infinite Index'' in IR}
\label{infinite-index-in-ir}

Figure~\ref{index-technologies} shows how we place the concept of an infinite index in the context of known information retrieval concepts. The basic and most widely used concept of an (inverted) index was defined by \citet{anderson:1997} as ``a systematic guide designed to indicate topics or features of documents or parts of documents.'' The topics or features of documents are represented by (index) terms. In modern information retrieval, these index terms correspond to the vocabulary of an indexed document collection. \mbox{\citet{anderson:1997}} further explains that ``[t]he function of an index is to provide users with an effective and systematic means for locating documentary units (complete documents or parts of documents) that are relevant to information needs or requests.'' Specifically, the documents that can be looked up in an index are stored elsewhere, with an index lookup providing the necessary information that identifies the storage location of the matching documents within the filing system.

This concept of indexing, invented long before the days of computers, is still used today, in the form of data structures that fulfill the definition and function of an index in the above sense. Most importantly, the inverted index data structure implements a mapping of index terms to so-called postlists, where each postlist is a list of ``postings'' containing, among other things, a document identifier for locating the document within a file system or document store. Recently, index data structures have been revisited in the context of research on neural information retrieval~\cite{mitra:2018,tonellotto:2022}: The neural index (the authors call it ``transformer-based generative indexing'')~\cite{bevilacqua:2022,tay:2022,wang:2022} has been proposed as a new type of index that mimics the function of a classical index by mapping queries directly to document identifiers. This mapping is trained based on a given document collection. Using an approach to predict queries that users might make to retrieve a given document, such as Doc2Query~\cite{frassetto:2019}, it is straightforward to generate training examples consisting of a triple of query, document, and the document's identifier, or even just tuples of identifiers and synthetic queries~\cite{zhuang:2022}. The goal of the model is to predict the identifiers of the relevant documents given a query.

In this paper, we propose a different way of indexing by using generative text-to-document models as indexes. Although we focus on images as documents, this type of indexing is in principle applicable to all types of documents. In this scenario, the ``index'' is trained using documents and texts describing the document as training examples. Unlike the indexing approaches mentioned above, the resulting model does not necessarily retrieve the documents that were part of the document collection used to train the generative model, but rather generates new documents. Thus, this indexing approach is different from the other two, while it can be considered as a kind of independent neural indexing approach.

\vskip1ex\noindent
Altogether, we classify the three indexing approaches as follows:
\begin{itemize}
\setlength{\itemsep}{1ex}
\item
Term-to-identifier indexing: building a lookup table that maps index terms to document identifiers.
\item
Query-to-identifier indexing: training a model to predict identifiers of relevant documents for a query.
\item
Query-to-document indexing: training a model to generate relevant documents for a query.
\end{itemize}
To find a technical name for these indexes, the following alternatives are suitable: ``generative index'', or ``neural index'', or, ``query-to-identi\-fier index'' vs. ``query-to-document index'', respectively.

\subsection{Interaction Methods}
\label{interaction-methods}

Although the characteristic way of interacting with generative text-to-image models is the text prompt, other features have been rapidly added to the interfaces to support the process of image generation. To illustrate the possibilities, we give here a brief snapshot of interaction methods based on the most common models (as of October~2022). Related generative text-to-video or text-to-3D models are not considered~\cite{ho:2022,poole:2022}.

\begin{figure*}
\begin{minipage}{.59\linewidth}
\centering\includegraphics[width=\linewidth]{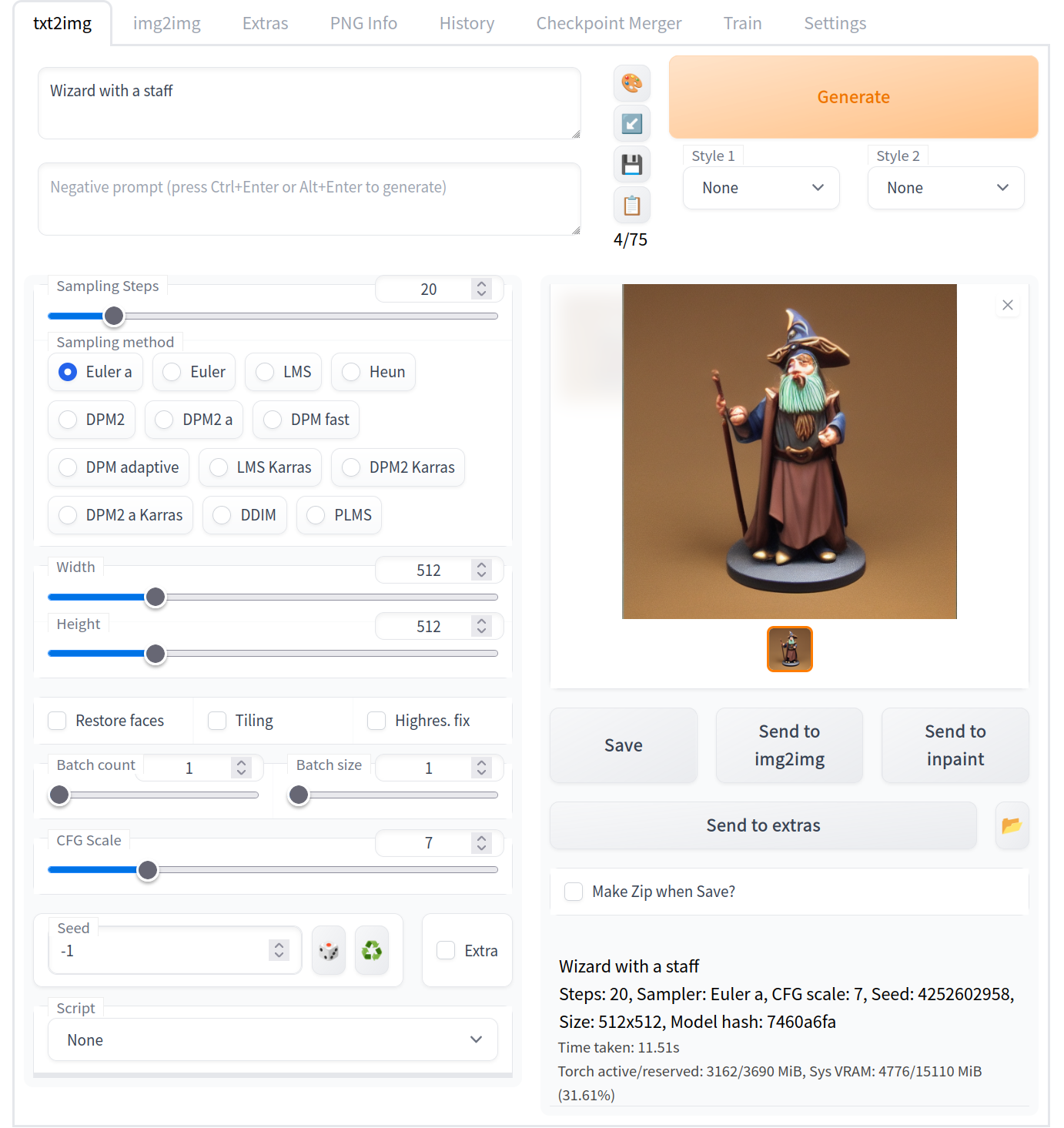}\\(a)\\[2ex]
\begin{minipage}{.48\linewidth}
\centering\includegraphics[width=\linewidth]{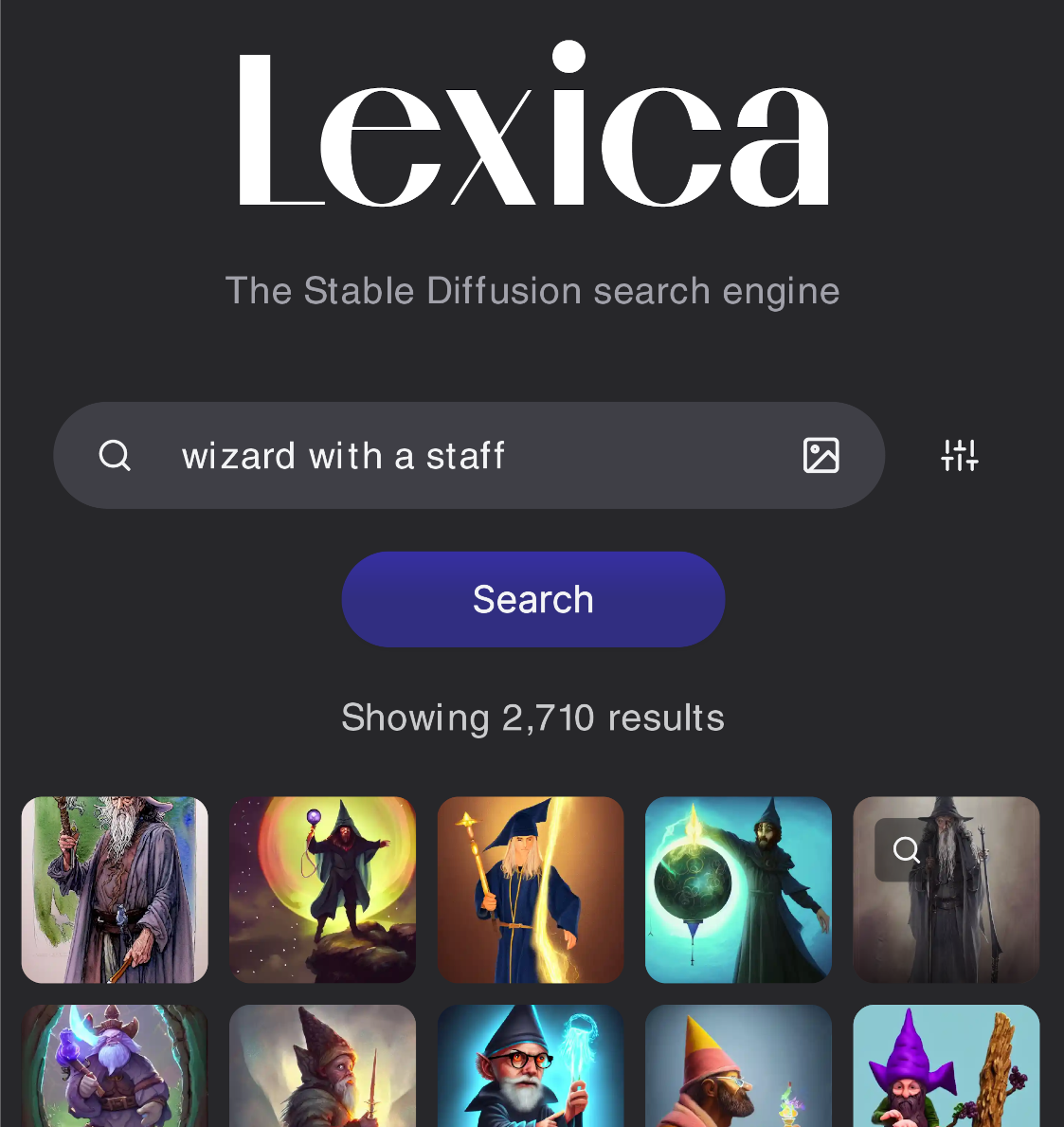}\\(b)
\end{minipage}%
\hfill%
\begin{minipage}{.48\linewidth}
\centering\includegraphics[width=\linewidth]{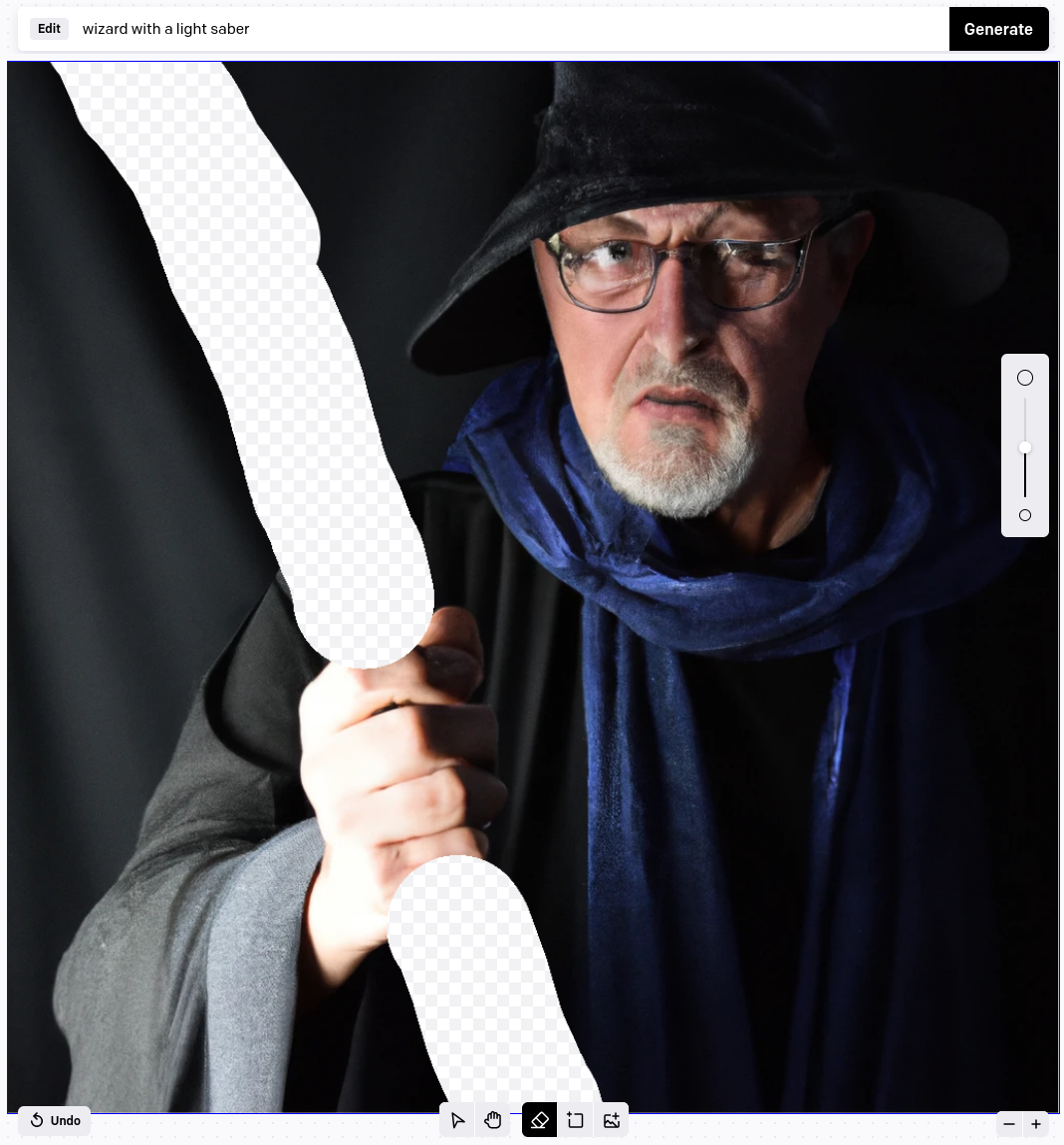}\\(c)
\end{minipage}%
\end{minipage}%
\hfill%
\begin{minipage}{.392\linewidth}
\centering\includegraphics[width=\linewidth]{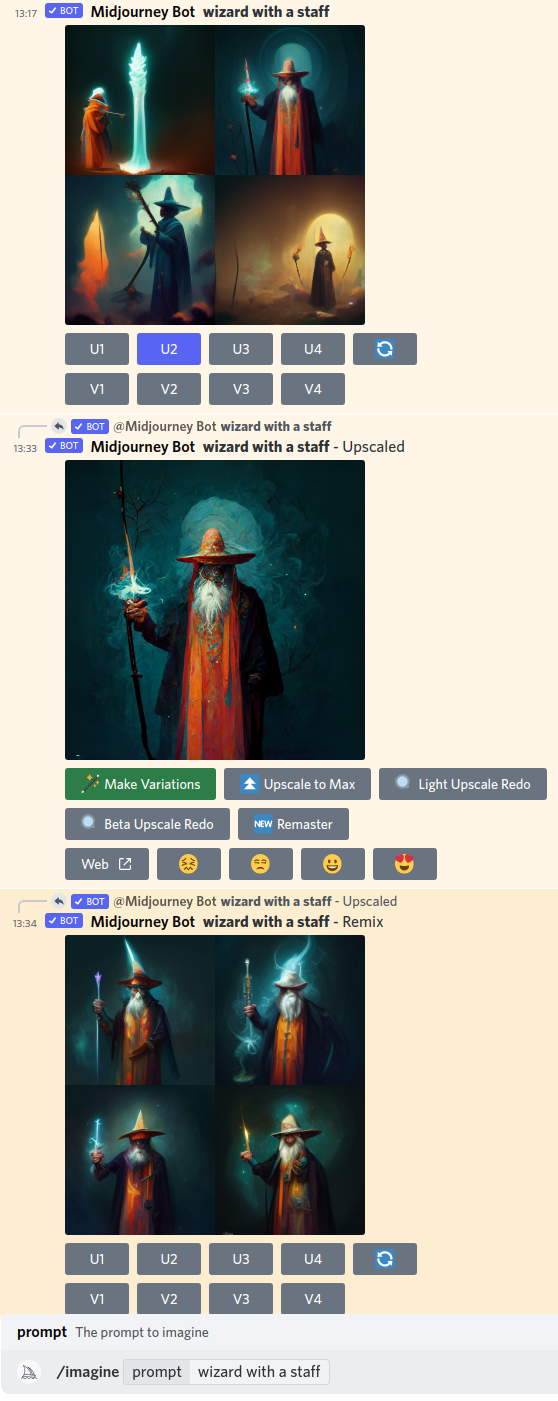}\\(d)
\end{minipage}
\caption{Screenshots illustrating the interfaces and interaction methods discussed in Section~\ref{interaction-methods}:
(a)~prompting in a community-maintained stable diffusion web interface;
(b)~Lexica search engine for generated images along their prompts;
(c)~in-painting in DALL$\cdot$E~2 on an image originally created for the ``Wizard with staff'' prompt: the staff was manually masked (shown in white) to produce a modified prompt;
(d)~upscaling and variation generation in Midjourney.}
\label{interaction-method-screenshots}
\Description{Four screenshots. Screenshot A shows a community-maintained Stable Diffusion web interface shows several text fields, sliders, and radio buttons, allowing to set prompt, negative prompt, sampling steps, sampling method, image width and height, face restoration, tiling, resolution fix, batch count and size, seed, and more. The output image is also shown, with buttons for saving or sending it to other interfaces for further AI-driven processing. Screenshot B shows the interface of the Lexica search engine. Below a Google-like search input it shows a grid of result images. Screenshot C shows a generated photo-like image of a bearded man in black wizard attire, holding his right hand as if he was holding a long staff in it, but the part where the staff would be is instead crudely erased, showing the almost white pattern used to depict transparency in image manipulation tools. Screenshot D shows a chat with three messages from "Midjourney Bot". The first message is named "wizard with a staff" and contains an image grid of four concept-art-style images depicting a wizard character with a staff. A button "U2" is highlighted below. The second message is named "wizard with a staff - upscaled" and contains a higher-resolution version of the second images in the previous grid. A button "Make Variations" is highlighted below. The third message is named "wizard with a staff - Remix" and contains a new grid of four wizard images, this time similar to the upscaled image from the previous message.}
\end{figure*}

\paragraph{Prompting}
For generative text-to-image models, prompting the model is the primary interaction method. This interaction method serves as the initial point of contact with the model during image generation, much like a query in a standard web search. The interaction method is identical for both: the user sends a short text and receives images in response. Some interfaces of generative models allow images to be included in the prompt to steer the generated images in a particular direction, much in the same way that content-based image retrieval is used to find similar images. Unlike content-based image retrieval, model interfaces typically require that the prompt also contains text. Another aspect of prompting in some interfaces is the specification of model parameters along with the prompt, e.g., the size of the image to be generated or whether to generate tiled images, which is similar to filters (e.g., by size) in regular web search. Moreover, the negation operator allows to exclude certain terms from the generated image. The widely used Stable Diffusion model provides only a command line interface, but the community has implemented several graphical interfaces for it, for example one maintained by AUTOMATIC1111%
\footnote{\url{https://github.com/AUTOMATIC1111/stable-diffusion-webui}}
(cf.~Figure~\ref{interaction-method-screenshots}a).

In addition, several services have emerged in the larger text-to-image model generation ecosystem to assist users with prompt engineering. Specialized search engines allow users to search for images created with generative text-to-image models. The search engines then reveal the prompts used to generate the images they find, allowing prompts to be reused. Images are indexed either by their prompt or by the image content (e.g., with~CLIP~\cite{radford:2021}). Examples of such search engines include the ``community feed'' of the Midjourney web app or the independent search engine Lexica, which indexes images from the Stable Diffusion Discord server (cf.~Figure~\ref{interaction-method-screenshots}b). According to the developer, 1.4~million queries were made in a week, the index contained 12~million images in September~2022, and 5~million USD was earned, which clearly indicates the need for such systems. Other services enable (social) prompt engineering in a click interface%
\footnote{E.g., \url{https://phraser.tech}}
or even to buy prompts that supposedly provide consistent results.%
\footnote{\url{https://promptbase.com}}
Other projects carefully analyze how the prompt affects the result and create extensive lists of examples.%
\footnote{E.g., \url{https://github.com/willwulfken/MidJourney-Styles-and-Keywords-Reference}}
Although these services have a similar goal as query suggestions in web search, namely to help with prompt engineering, their interaction pattern is different. We discuss the implications in Section~\ref{part5}.

\paragraph{Variations}
When generating an image, the variations interaction method allows to change parts of the image composition. This is useful when a generated image is broadly satisfactory but needs improvement in certain aspects. We distinguish three ways of generating variations:
\Ni
the user does not change the prompt, which causes the composition to change only slightly and randomly (cf.~Figure~\ref{interaction-method-screenshots}d);
\Nii
the user changes the prompt and gives the model a new target as it continues from a generation checkpoint of the original image;
\Niii
the user specifies semantic processing of the image, changing elements of the original image while preserving its original characteristics~\cite{kawar:2022}. 
This interaction method, especially in the case of~(1), is similar to the ``show similar results'' button in regular image retrieval. However, (2)~and especially~(3) allow a clearer specification of the need.

\paragraph{In- and Out-Painting}
When generating an image, in- and out-painting allows to limit the generation of variations to user-defined areas of the image. This is useful when the user wants to change a certain area of the generated image (in-painting; cf.~Figure~\ref{interaction-method-screenshots}c) or expand an image (out-painting), where the model tries to fill the region to match both the prompt and the parts of the original image at the edge of the region.  This interaction method goes beyond the capabilities of regular search interfaces, and in most cases one would expect finite indexes to contain no matching results. For an infinite index, this interaction method can be extremely useful to finding images that satisfy multiple requirements.

\paragraph{Quality Enhancements}
If the user is satisfied with the composition of an image, quality enhancement allows improving the image quality in one or more ways without changing the composition. The most common way to improve quality is to upscale the image to a higher resolution. There are often various upscaling methods that create new versions from a source image that look sharp or soft, realistic or artistic, without losing the original composition. Choosing a specific upscaling algorithm is useful to generate different images that should look similar in terms of their composition. Another type of enhancement is the use of image-to-image models trained specifically for correcting faces~\cite{wang:2021}. We anticipate that other image-to-image models specializing in specific operations will be integrated in the future. As with the variations tool, the closest counterpart to this method in regular search is the ``show similar results'' function, which can be quite effective for finding higher resolution images. However, quality enhancements allow a much clearer specification of what is needed by comparison.

\paragraph{Image-to-Text}
If the user wants to rephrase the prompt but also use parts of the generated images, image-to-text models can be used to obtain a textual description of the image that reads like a prompt. We are not yet aware of any regular image search engine that integrates image-to-image models on the user page, although we believe that major image search engines such as Google Images will use them to index images.

\subsection{Relevance in Text-To-Image Generation}
\label{missing-intermediary}

As the above overview of interaction methods shows, the text-to-image generation community develops support for a variety of common search problems, but so far used information retrieval concepts only as search facets supported by external tools. This section reviews relevance as a core information retrieval concept that needs to be operationalized to steer the generation.

As with regular image retrieval, also generative models the concept of result relevance depends on the information needs of users, of which seven different categories have been identified in the literature (cf.~Section~\ref{part2}). Generating images rather than finding them can, at least in theory, satisfy most of these needs, and is particularly useful for the needs of entertainment, illustration, aesthetic appreciation, engaging others, inspiration, and social interaction. In social interactions, for example, it is very useful for generative models to take into account the general moods mentioned in the prompt, providing a clear path to generating images that evoke specific emotions. The one information need category for which image generation is unsuitable is the need for knowledge construction, since generated images are not tied to real world knowledge.

When generating images, a distinction must be made between two different intentions. First, the user may already have a clear idea of the target image, for example, in an illustration. A user with this intention iteratively refines their prompt until the system generates an image that approximates their ideas, which we call a {\em descriptive approach}. Second, they may not have a clear vision or goal, just a set of constraints. With this intent, the user iteratively refines their prompt in a feedback-loop with random elements introduced by the system, loosely steering the system toward an image that they like and that meets the constraints, which we call the {\em creative approach}. Although they are very different from the user's point of view, both approaches are more or less indistinguishable for the system in terms of query log analysis: a general prompt is extended with details to become more specific.

With respect to text-to-image model-based retrieval, the research in interactive information retrieval is highly related (cf.~Section~\ref{part2}). Query log analysis will be important to identify keywords in prompts that generally produce satisfactory results, to model user intent at a finer level, and to identify search queries and early abandonments that may indicate problems in the model. We assume that query suggestion methods will be very helpful, especially to assist inexperienced users. However, automatic query reformulation for prompts is more challenging because such changes have a generally more unpredictable impact on the generated images. In our case study (Section~\ref{part4}), the creative professional therefore refrained from optimizing the prompt and instead tried completely new ones. We see here a clear lack of user support in terms of retrieval in the current interfaces. External tools such as prompt search engines attempt to compensate for this shortcoming, but cannot match the effectiveness of integrated solutions that are widely used in search engines today (see Section~\ref{part5} for a discussion of possible remedies).

With these considerations in mind, the notion of relevance and thus retrieval methods such as query suggestions can be transferred from information retrieval to text-to-image model generation, and thus retrieval evaluation measures can be adopted.

\subsection{\fontsize{10.5pt}{9pt}\selectfont Evaluating Retrieval on Text-To-Image Models}
\label{ir-evaluation}

Framing text-to-image generation as a retrieval problem implies measuring the effectiveness of generated rankings of generated images according to standard experimentation practices in information retrieval. However, we show that the infinite index in the form of a text-to-image model has far-reaching consequences for the design and evaluation of experiments, since the set of relevant documents is not closed and can thus not form the basis to calculate recall. We also discuss the challenges this poses for creating reusable benchmark collections and speculate on approaches to overcome these challenges. We focus on measuring ranking effectiveness because other aspects, such as user interface design and layout, are not considered in Cranfield-style evaluations.

\paragraph{Impact of the Infinite Index on IR Evaluation Measures}
Effectiveness measures can be divided into utility-oriented (based on a ranking only) and recall-oriented (normalized by a ``best possible'' ranking) evaluation measures~\cite{lu:2016} so that an appropriate measure can be used depending on the nature of the information need. However, the virtually unending stream of alternative images that can be generated leads to problems with recall-oriented evaluation measures. An infinite number of images that can be generated allows for the subset of highly relevant images is to also be infinite. For recall-oriented measures like nDCG~\cite{jarvelin:2002}, this means that their normalization term can default to a ranking that is completely filled with highly relevant images. In practice, a human will still only search a query up to a certain rank~$k$, so an nDCG@$k$ can still be computed in this way, since a specific retrieval model requesting a text-to-image model may still deviate more or less from actually providing only highly relevant images. Utility-based measures (such as Precision@k, MRR, RBP~\cite{moffat:2008}, etc.) are not affected by this problem because they measure the effectiveness of a ranking based only on the images available in the ranking. 

Another problem is that an infinite number of near-duplicate images of high relevance can be generated. Retrieval models could therefore rank many/exclusively (near-)duplicate images highly. If evaluated in isolation, each one would be considered highly relevant. Evaluation measures that operate on rankings with (near-)duplicates overestimate their effectiveness~\cite{bernstein:2005,froebe:2020a}, and learning-to-rank approaches learn suboptimal ranking models as well when trained on redundant data~\cite{froebe:2020c}. Therefore, it is important to deduplicate the rankings before evaluation. For the development of retrieval models, this means that ensuring diversity of images in the top ranks can be instrumental for users.

Overall, utility-based measures (such as~RBP) on deduplicated rankings with judgments for the top-$k$ images allow theoretically grounded evaluations when using text-to-image models as index.

\paragraph{Evaluations with Active Judgment Rounds}
Experimental evaluation of retrieval systems usually follows the Cranfield paradigm~\cite{cleverdon:1967,cleverdon:1991}, which assumes that all documents are judged for all information needs. The original Cranfield experiments~\cite{cleverdon:1967,cleverdon:1991} were conducted on a collection of 1,400~documents and complete relevance judgments for 225~topics. However, complete judgments became impracticable almost immediately thereafter as the size of collections increased significantly. The current best practice for shared tasks in~IR is to create pools of the top-ranked documents from the submitted systems for each topic and then score each topic's pool~\cite{voorhees:2019}, assuming that unjudged documents are not relevant. However, the assumption that judgment pools are ``essentially complete'' is likely incorrect when text-to-image models are used as index, especially if query expansion approaches are involved. As a result, rigorous evaluations must include manual rounds of judgments of unjudged images to reestablish ``completeness'' (e.g., for the top-$k$ results), at least for utility-oriented measures, which hinders fully automated evaluations.

\paragraph{Evaluations without Active Judgment Rounds}
IR research has benefited largely from the availability of robust and reusable test collections created during shared tasks~\cite{voorhees:2001}. However, these collections are robust only if most of the unjudged documents are irrelevant, which is not the case for text-to-image models. Consequently, creating robust and reusable test collections is a major challenge that requires experience from several different shared tasks and subsequent post hoc experimentation (e.g., some robustness checks for traditional test collections are not performed until years after their creation~\cite{voorhees:2021}). Therefore, any post-hoc experiments based on an infinite index would need to include appropriate handling of unjudged images. Traditionally, unjudged documents are either simply removed (where a system's result lists are condensed to the included judged documents in their relative order)~\cite{sakai:2007a}, classified as not relevant (default setting) or highly relevant (lower/upper bound)~\cite{lu:2016}, or their relevance labeling can be predicted~\cite{aslam:2007}. While these approaches are well studied for conventional retrieval experiments (e.g., condensed lists often overestimate the effectiveness values~\cite{sakai:2008} and the gap between lower and upper bounds can be very large~\cite{lu:2016}), it is not yet clear whether they are suitable for an infinite index. As a result, it is not yet clear how to construct robust and reusable test collections, but we speculate that techniques from machine translation (e.g., measuring the similarity of an unjudged document via phash~\cite{zauner:2010} to judged reference images) or relevance prediction may be appropriate.

\subsection{The First Index for The Library of Babel?}

At the beginning of the 20th century, Kurd La\ss{}witz, a German writer, scientist, and philosopher who became the first German science fiction author, introduced ``The Universal Library''~\cite{lasswitz:1897} as part of a series of short stories published in a newspaper around that time. The Universal Library contains every conceivable book with a length of 1~million characters. Assuming an alphabet of 100~Latin letters, numerals, and punctuation marks, each combination of these characters in a book of 1~million characters yields $10^{2,000,000}$~books, virtually everything that can been written in every language (assuming an appropriate transliteration). The only problem with such a library is that it is extremely unlikely to find a book by chance that contains a plausible sentence. This idea was taken up by Jorge Luis Borges, a well-known Argentine author, and made widely known under the name ``The Library of Babel''~\cite{borges:1941}. He imagines this library as a universe of its own and invents stories about various tribes of humanity that might develop in such a place, always looking for scraps of knowledge among the many books of incomprehensible gibberish. In an earlier work called ``The Total Library''~\cite{borges:1939}, Borges traces the history of this concept back to La\ss{}witz and even to Aristotle and Cicero, who formulated what is now known as the ``Infinite Monkey Theorem''~\cite{wikipedia:infinite-monkey-theorem}, which states that a monkey hitting a typewriter at random will eventually type every text, including the complete works of William Shakespeare.

Given this fictional concept, generative text-to-document models can be understood as an index and a search engine for the library of Babel: By entering a short phrase as a query, the model is prompted to search the library for a document that matches the query. This completely circumvents the problem outlined by La\ss{}witz and Borges, since a document returned by a generative text-to-document model is very likely to be related to the query, and as long as the query itself is not gibberish, the retrieved documents will not be gibberish either. 


%% file: chiir23-generative-model-as-index-part4.tex
\section{Case Study: Game Artwork Search}
\label{part4}

To illustrate retrieval using a query-to-document index for images, we report on an observational case study in which a text-to-image model is used for a creative task. First, we describe the study setup and the exemplary creative task, generating graphics for an online card game (Section~\ref{case-study-setup}). Subsequently, the main observations of the study are summarized (Section~\ref{case-study-summary}). A full report on the study is available as supplementary material.%
\footnote{Case study report: \url{https://doi.org/10.5281/zenodo.7221434}}

\subsection{Setup of the Case Study}
\label{case-study-setup}

For the case study, we recruited a creative professional through personal contacts who allowed us to observe him as he explored the use of generative text-to-image models in his creative process. The professional described himself as a game designer and developer with the experience of five major game releases and as a lecturer in game development at a university. Prior to the case study, he described himself as very intrigued by generative text-to-image models he had come across in his Twitter feed, and had also seen some online videos on this technology (``2~minute papers''). Moreover, he had already generated about 50~images in DALL$\cdot$E~2, about 20~in Midjourney, and less than 10~with Stable Diffusion on his own hardware, but none of them as part of a project. He anticipated, however, that generative text-to-image models will become very useful for the video game industry.%
\footnote{Video games account for about~57\% of digital media market revenue in~2022, or US\$197~billion~\cite{statista:2022}. Meanwhile, other game developers have also published reports on their experiments with text-to-image models, e.g., \url{https://www.traffickinggame.com/ai-assisted-graphics/}}

Based on his experience, the professional decided to investigate the use of generative text-to-image models in the creation of graphics for an online card game for the study. Specifically, he was interested in developing a ``deck-building online card game like Magic the Gathering set in a fantasy universe.'' In this game, each playing card has its own artwork that visually links it to the fantasy universe. Moreover, the cards belong to different ``factions'' that must be visually distinguishable. The professional opted for a ``concept art-like style'' from the outset. In the five hours we provided for the study, the professional expected to first create a ``mood board'' of images to capture the artistic style of the desired artwork~\cite{lemarchand:2021}, and then create the artwork itself for some cards. Based on his own testing, he decided to use Midjourney for this task. This choice reflects Midjourney's concept, which emphasizes ``painterly aesthetics'' and aims to help creatives ``converge on the idea they want much more quickly''~\cite{salkowitz:2022}, especially at the beginning of a project.

The case study was conducted using the think-aloud method, asking additional questions while the professional waited for the images to be generated. Since the study did not focus on search interface design, one of the authors used Midjourney extensively to prepare for the study and provide technical support to the professional. To record observations, we took extensive notes as well as video and audio recordings and used the logging capabilities of the Midjourney web app. Following~\citet{li:2022}, we used forms to structure our notes for various events, in our case for queries, problems, and shifts in design goals. A report on the study with all generated images is available as supplementary material.

\begin{figure*}
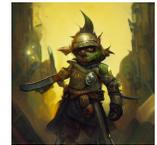
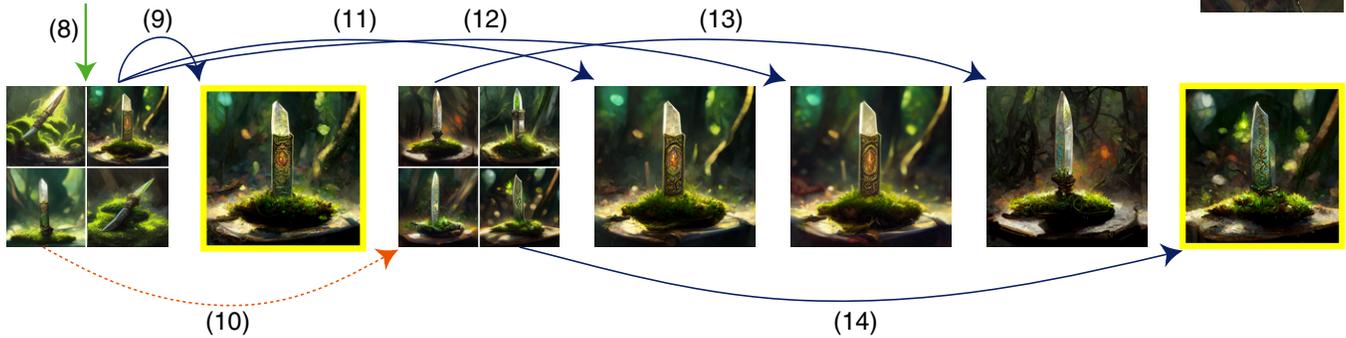

\includegraphics[width=\textwidth]{case-study-example-process}
\caption{Exemplified search for a generated image from the case study, consisting of 14~steps in 22~minutes. Gray prompt text is copied from the prompt of another image in the mood board. For the reformulated prompt---after the first series of images has been abandoned as ``leading nowhere''---, it is copied from the image that is part of the prompt. Interactions are text-to-image generation of four images (\raisebox{-1.5px}{\includegraphics{case-study-arrow-prompt}}), generating four variations of one image (same prompt, \raisebox{-1.5px}{\includegraphics{case-study-arrow-variations}}), and upscaling one image (\raisebox{-1.5px}{\includegraphics{case-study-arrow-upscales}}). The ``beta'' upscaling method is used in Step~12, the ``light'' method in Step~13, and the default method (``detailed'') in all other cases. The professional kept the two images with the yellow border. Although the image generated in Step~9 did not show a dagger as intended, he found it intriguing and said that it evoked a story, especially in combination with the kept image.}
\label{case-study-example-process}
\Description{A sequence of 14 generated images and image grids, broken into two parts. The first part contains 7 images and image grids generated from the following initial prompt with quality set to 2: Prompt start. An ancient golden dagger lying on moss, illuminated by godrays, close up, digital painting, matte painting, midjourney, concept art, detailed art, scifiart cinematic painting, magic the Gathering, volumetric light, masterpiece, volumetric realistic render, epic scene, 8k, post-production detailed art, scifiart cinematic painting. Prompt end. Image grid 1 shows four images of 1 or two golden daggers with artistic ornaments on green moss. Image grids 2 to 5 are generated variations of the image created for this prompt. Images 6 and 7 are generated high-resolution versions of one image from the image grid 5. The second part contains another 7 images and image grids. Image grid 8 is generated from a reformulated prompt, for which the part "an ancient golden dagger lying on moss, illuminated by godrays, close up" is changed to "an ancient golden dagger lying on moss, illuminated by godrays, close up". This time, only image grid 10 is a variation of image 8. The 5 remaining images are upscaled versions from the two image grids. Two images are marked as being selected by the professional for their task. One of these images looks more like depicting a crystal on a some artistic tube than a dagger. In the other image, the blade of the dagger sticks out from the moss, with the handle not visible.}
\end{figure*}

\subsection{Main Observations from the Case Study}
\label{case-study-summary}

This section summarizes the insights from the case study into three main observations. We found that the mood board is a key tool for professionals and analyze its use based on the five reasons for using information resources~\cite{li:2022}. To analyze the mental state of the professional, we use \citeauthor{kuhlthau:1993}'s~\cite{kuhlthau:1993} model of the information search process. And based on the professional's comments during the study, we identified the lack of control he mentioned as the main problem that needs to be addressed by future tools.

\paragraph{The mood board as prompt library}
\citet{lemarchand:2021} defines a mood board as ``a single page or screen of pictures arranged around a certain idea or theme'' that serve two main purposes: first, to inspire new ideas by juxtaposing images (supporting creative processes), and second, to communicate a concept quickly and effectively (managing found information). After creating the mood board from images in Midjourney's community feed, however, the professional immediately began using the mood board as a source for his prompts as well. When creating a new image, he selected from the mood board the image that came closest to his ideas in terms of artistic style, and then copied the ``style part'' of that image's prompt for his own creation (cf.~the gray text in Figure~\ref{case-study-example-process}).
\enlargethispage{2\baselineskip}
Thus, he additionally used the mood board to learn domain knowledge (style names, rendering engines, etc.) and procedural knowledge (parameters such as \mbox{\tt``-{}-q~2''} to increase image quality). Only once did the professional search for the artists of the ``Magic the Gathering'' cards using an external search engine and was pleased to find that they were already included in the prompts he copied. Learning happened only on a superficial level, copying entire style sections of a prompt and using it like an atomic unit. This behavior is so widespread in the text-to-image generation community at the moment that commercial services have emerged for them.%
\footnote{E.g., \url{https://promptbase.com}}

\paragraph{Uncertainty never fully ceases}
In \citeauthor{kuhlthau:1993}'s~\cite{kuhlthau:1993} model of the information-seeking process, the seeker moves from uncertainty to understanding as the search progresses. During the case study, we were able to identify clear parallels to this model and its phases, particularly the selection, exploration, formulation, and collection phases. In the selection phase, the professional uses the mood board as inspiration to choose content and style for a new image. In the exploration phase, he created and modified the prompt: he mentioned that he was very unsure about the results he would get and how he could modify the prompt to achieve what he envisioned. Once he found something he thought was promising, he moved into the formulation phase, focusing on generating variations over and over again and figuring out certain aspects that the final image should have. With a clear sense of direction, he would then upscale matching images in the acquisition phase and test the various upscaling algorithms as necessary. As accounted for in the model, the professional also regressed to earlier stages, especially when he saw an impasse (cf.~Figure~\ref{case-study-example-process}). \citeauthor{kuhlthau:1993}, however, mentions two ``types of uncertainty,'' and although uncertainty about the concept (what he is looking for) decreases as described above, uncertainty about the technical process (how to get there) remains high, with the AI remaining largely unpredictable to him.

\paragraph{Sense of direction, but lack of control}
Although in some situations the professional noted that the unpredictability inherent in the process was appealing (``I also wanted to be surprised''), he also mentioned that the process was very exhausting, which we related to the fact that he often went back in the history of his generated images to keep checking which interactions yielded good results and which image he should continue with. An interface that supports the user in organizing generated images therefore seems necessary. The professional noted that he was developing a sense of the direction the image variations would take, but also felt he had no control. He decided whether to continue down one path or try another, but did not feel he could change direction. After the case study, Midjourney introduced the ability to modify a prompt when generating variations, but the professional says this does not solve the problem of choosing the right words. Uncertainty about how to change the prompt to achieve the desired results therefore has a major negative impact on the user's sense of control.

\vskip2ex\noindent
Indeed, the case study showed clear parallels between text-to-image generation and image search. In particular, we found that existing theoretical models of the (creative) search process are broadly applicable. The main difference lies in the never-ending uncertainty about how to get to a particular result---although the user must assume, because of the index being virtually infinite, that there is a path that leads to the goal. Based on our observations, we believe that tools that provide the user with more intuitive ways to control the generation process are needed to bridge this gap.

%% file: chiir23-generative-model-as-index-part5.tex
\section{Discussion}
\label{part5}

Based on our conceptualization of text-to-image models as search and the case study, we next explore the limitations of text-to-image generation (Section~\ref{sec:limitations}). Then we discuss how active learning might help (Section~\ref{sec:active-learning}), and address ethical concerns (Section~\ref{ethical-concerns}).

\subsection{Limitations of Text-To-Image Generation}
\label{sec:limitations}

While the functionality of text-to-image models is already of sufficient quality to be used in real-world applications~\cite{dalleblog:2021,salkowitz:2022}, we identified the following two main limitations related to the workflow or capabilities of the current methods---the same workflow that was used in the case study.

\paragraph{Prompt Engineering}
Although prompt engineering has been successfully applied to other generative tasks such as co-writing screenplays and theater scripts with a large language model~\cite{mirowski:2022}, the need to engineer the prompt compromises the intuitiveness of the prompt interface. Users quickly realized that iteratively adding modifiers to the prompt (as in Section~\ref{part4}), causing the model to apply the desired result styles to the generated image, is the most effective way to control the image generation process~\cite{liu:2022,oppenlaender:2022}. This has given rise to a whole new subfield of text-to-image prompt engineering~\cite{liu:2022}, where prompts are increasingly becoming long strings of keywords instead of text descriptions. These manipulated prompts resemble highly optimized search engine queries where users select and fill in keywords---so users have learned to adapt to the algorithm rather than the other way around.

\paragraph{Influence of the Training Data}
A fundamental limitation of current models is that both text encoders and diffusion models generate new data by merging concepts learned from large datasets and are thus limited to those concepts. Writing a prompt that contains a concept that does not appear in either the text corpora or the image datasets is likely to result in sub-par generation of images. One possible remedy is that unknown terms can be described as paraphrases. If the training data does not contain images of a centaur, a prompt such as ``{\tt\small a mythical creature with the body of a horse and the torso of a human}'' might still produce the desired result.

\subsection{\fontsize{10.5pt}{11pt}\selectfont Active Learning for Text-To-Image Generation}
\label{sec:active-learning}

From the case study, it appears that targeted text-to-image generation is already surprisingly effective. As described in Section~\ref{interaction-methods}, the current way of working amounts to iterative prompt engineering, which in turn is a fundamental limitation, as stated in Section~\ref{sec:limitations}. We propose active learning as a solution to this problem and outline how it can be integrated as a feedback mechanism in an image generation workflow that uses text-to-image models.

\enlargethispage{\baselineskip}
Active learning~\cite{lewis:1994,zhang:2002} is an iterative approach to classification that involves a feedback loop involving a user and a (semi-)super\-vised machine learning model. It is intended for scenarios where training data is not available to minimize the effort required to obtain a suitable labeled training dataset while maximizing model quality.
According to \citet{schohn:2000}, an active learning setting consists of
\Ni
a model that is trained for a specific task,
\Nii
a query strategy that selects data from an existing resource or generates new data to be labeled, and
\Niii
a stopping criterion that indicates at what point continuing the process is unlikely to sufficiently improve the result any further.
At each iteration, the query strategy selects the examples it deems most informative for the model, for example, based on the prediction uncertainty of the model~\cite{schroeder:2022}. These examples are then annotated by the user according to the task at hand. A new model is then trained on all previously marked data, and the loop is repeated until an objective stopping criterion is met or the user stops. 

\begin{figure}[t]
\centering
\includegraphics[width=\columnwidth]{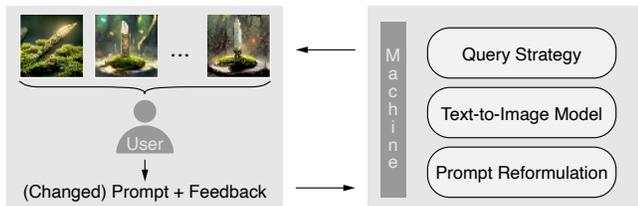}
\caption{A conceptual overview of the active learning loop for the guided text-to-image generation use case.}
\label{active-learning}
\Description{Two boxes that provide a conceptual overview of the active learning loop. Inside the left box there is ab active learning batch depicted, illustrated by three single images, each of which shows a variant of the golden dagger from the previous Figure. These images are then shown to a user, represented an icon depicting a human, who can subsequently provide feedback for the given images, and who can also reformulate the prompt in response the presented images. Next, on the right edge slightly outside the left box, an arrow leads from the left box to the right box. The right box shows the algorithmic active learning part, which is captioned as "machine" on the side. In this part two rounded boxes are shown: text-to-image model and query strategy. The text-to-image model, will likely include or be preceded by a prompt reformulation step. After this has incorporated the updated user feedback, new images are generated, from which the query strategy then selects a batch to be shown to the user. An arrow slightly outside the right box leads back from the right box to the previous box on the left, completing the active learning loop's cyclic structure.}
\end{figure}

For text-to-image generation, the whole structure of active learning is shown in Figure~\ref{active-learning}. The process begins with the user and an initial prompt. The active learning model learns to reformulate prompts, which in turn are passed to the text-to-image model. The model is trained with user feedback as target values, so that the resulting images should become increasingly appealing to the user. Subsequently, the query strategy decides which images are displayed to the user. It strikes a balance between exploration and exploitation, a well-known trade-off in information retrieval: exploration selects images that are different from the current best candidates, and exploitation selects images that are close to the current best solutions. Finally, the stopping criterion is the user who stops the process as soon as his information need is satisfied. In this setup, active learning uses relevance feedback~\cite{tong:2001,zhang:2002,xu:2007}.

Information retrieval systems can let users explicitly specify relevant documents (explicit relevance feedback) or learn from passive observations (implicit relevance feedback)~\cite{white:2005}, though this discussion focuses on explicit feedback to guide active learning for image retrieval. There are different types of explicit relevance feedback for the user:
\Ni
binary relevance feedback~\cite{gay:2009}, where the user rates each image as ``unappealing'' or ``appealing'' with respect to the target concept;
\Nii
graded relevance feedback~\cite{gay:2009}, in which the user rates each image from ``unappealing'' to ``appealing'' on a multilevel scale (e.g., from~0 to~5);
\Niii
ranking, where the user rates each image (possibly including images from previous iterations) from unappealing to appealing.
Users can provide feedback on the entire image or on individual parts (e.g., the background) or aspects (e.g., the color scheme). Similar to query customization during a regular search, the user can change the prompt in each iteration.

The main challenge for this feedback mechanism is to convert the images into a textual representation that preserves the specifics of each image, which can then be used to learn how to reformulate the prompt. For example, a prompt like ``{\tt\small wizard with staff}'' could generate images with different poses and backgrounds. To learn reformulations from relevance feedback, it is necessary to obtain a textual representation that includes these differences. One could, of course, try to learn to reformulate based only on latent image representation and relevance feedback, but this would solve the problem exclusively in the image space and largely ignore the text embedding space. This could also be a useful approach, but is outside the realm of natural language processing and information retrieval. Although the reverse step of image-to-image generation required for this has recently attracted increasing attention~\cite{rinon:2022,gal:2022}, it remains a challenge, and moreover, multiple images are required to generate one text~\cite{gal:2022}. Once this reverse direction is improved, the full spectrum of natural language processing and information retrieval can be applied to effectively process user feedback to improve prompts during the reformulation step.

When text-to-image generation is viewed as a retrieval problem (as in Section~\ref{part3}), the process of trying different prompts until a satisfactory image is generated is similar to traditional image retrieval, and thus the inclusion of active learning as a relevance feedback mechanism is an obvious choice of a well-established method. We anticipate that active prompt generation will be a strong interface competitor for generative text-to-image models once image-to-text models are sufficiently mature (apart from editing options such as in-painting or out-painting, which are orthogonal to this approach). 

\subsection{Ethical Concerns}
\label{ethical-concerns}

A computational approach powerful enough to generate documents such as images, text, and other media types at a quality difficult to distinguish at times from human-made illustrations naturally raises ethical concerns. We discuss the most important ones below.

\paragraph{Will algorithms replace artists?}
We begin with the obvious question: will generative text-to-image models threaten artists' jobs? First, based on our experience in the case study, it is currently difficult to get text-to-image models to generate a desired result. The decision whether the generated images represent the desired scene with sufficient quality still has to be made by the user. Therefore, we believe that these new models will be a powerful tool, but will not replace the human illustrator in the foreseeable future---even if the image quality should eventually reach human levels. This is corroborated by others such as \citet{liu:2022b}, who developed and evaluated a system that assists users in generating images for news articles, noting that artistic knowledge is still beneficial to the generated result, explicitly saying ``generative AI deployment should [...] augment rather than [...] replace human creative expertise''. We support this view: instead of an autonomous~AI that acts on its own, we want to emphasize the benefits of a ``supportive~AI'' that inquires about and incorporates the decisions of its users.

\paragraph{Who is the author of a generated image? And who owns the rights?}
This is currently an unresolved situation that leads to uncertainties regarding the use of AI-generated images. For this reason, major platforms such as the well-known image provider Getty Images have recently banned all AI-generated content.%
\footnote{\url{https://voicebot.ai/2022/09/23/getty-images-removes-and-bans-ai-generated-art/}}
Stakeholders may include the user, the creators, and the artists who created the images used for model training. Ultimately, this decision must be made by policy makers and by the courts, where many legal precedents have been set in the past through copyright litigation.

\enlargethispage{\baselineskip}
\paragraph{Text-to-image models for generating misinformation?}
Generated misinformation is already a pervasive problem and is widely discussed in the context of so-called ``deep fakes'' and AI-generated text~\cite{zellers:2019,schuster:2020,kreps:2022}. To mitigate this problem in text-to-image models such as Stable Diffusion, an image is watermarked to identify it as artificially generated.%
\footnote{\url{https://github.com/CompVis/stable-diffusion}}
Although watermarks are not easy to remove, this may not be enough if they are not checked on virtually all devices. However, this requires that policymakers legally oblige device manufacturers to detect fakes and warn users. In addition, watermarking images itself raises privacy concerns. As for text-to-text models, fully generated documents can be useful, provided they are not used to generate factual knowledge, which is currently woefully inadequate. Therefore, the use of such models as an infinite index must at least be subjected to post-processing in the form of fact checking or the like.
This is exactly what is happening at present, after OpenAI recently introduced ChatGPT%
\footnote{\url{https://openai.com/blog/chatgpt/}}
with lots of publicity: The search engines You%
\footnote{\url{https://blog.you.com/a9e05080c8ea}}
and Neeva%
\footnote{\url{https://neeva.com/blog/introducing-neevaai}}
have already integrated facsimiles of ChatGPT into their search interfaces and check the generated documents against traditional search results. Whether this proves to be a good idea remains to be seen.

\paragraph{Do these models express or even amplify bias?}
Bias in training data is a known problem for both image data~\cite{khosla:2012} and language models~\cite{kirk:2021}. Therefore, text-to-image models must also be systematically screened for social and other types of bias. In information retrieval, for example, fair ranking is now a widely studied problem. A retrieval process built on generative models could be designed to mitigate their inherent biases. Image search engines based on generative models must post-process and re-rank their results to compensate for bias, just like their traditional counterparts. However, the technologies developed for traditional search engines can also be applied to search engines based on generative models.%
\footnote{For example, compare the result from lexica.art (\url{https://lexica.art/?q=nurse}) with that from Google Images (\url{https://www.google.com/search?tbm=isch&q=nurse}).}

%% file: chiir23-generative-model-as-index-sum.tex
\section{Conclusion}
\label{conclusion}

\begin{figure}
\includegraphics[width=.475\linewidth]{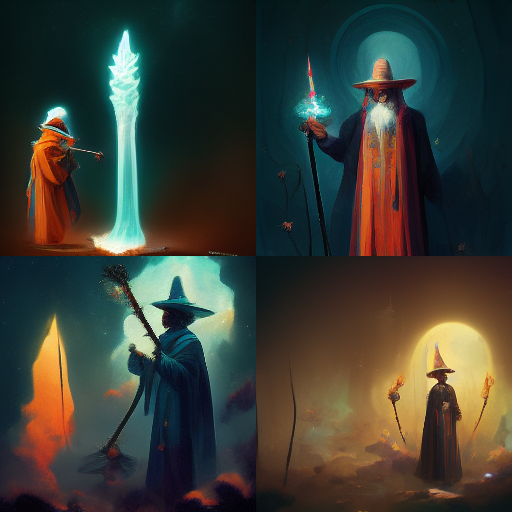}%
\hfill%
\includegraphics[width=.475\linewidth]{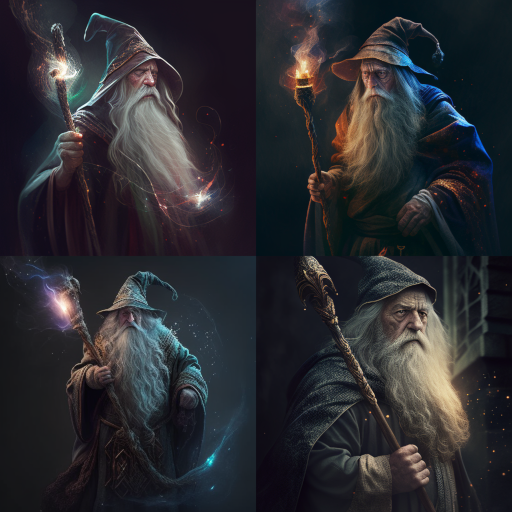}
\caption{Results for the prompt ``wizard with a staff'' in Midjourney: (left) version 3, default at the time of our case study; (right) version 4, the default three months later.}
\label{midjourney-wizard-with-a-staff-comparison}
\Description{Two grids of four images each. Each image shows a bearded male character with a wizard's hat and a long staff in his hand, that in some images has a magical glow. The images on the left, from Midjourney version 3, look like concept art, whereas the images on the right, from Midjourney 4, look like almost realistic portraits.}
\end{figure}

Supporting systems and services are needed for the use of generative text-to-image models. Their integration into existing systems is already in full swing, as has been seen for years in generative models for writing assistance and translation systems, but now also in more creative areas. However, integration with end-user software to create slide presentations or artwork will not meet all the needs of those looking for inspirational images. Given the recent moves by You and Neeva, specialized search engines based on generative text-to-image models as indexes, with user interface for formulating information needs and customized retrieval models, are probably already being developed. However, the development of a search engine is not trivial, and the information retrieval community faces the renewed challenge of developing an understanding and technological foundation for such search engines. This includes the development of new retrieval models and relevance scores as well as the adaptation of evaluation methods for benchmarking search engines based on generative models. Moreover, because results can vary widely from one day to the next (cf.~Figure~\ref{midjourney-wizard-with-a-staff-comparison}), users cannot rely on things like remembering specific queries to search for known items. Therefore, to effectively use generative image models as a search index, it may be necessary to maintain a history of search results with appropriate model parameters. Finally, what is true for generative text-to-image models is likely to be true other kinds of text-to-document models, opening up a whole new world of exciting new research directions and promising high impact.

